\documentclass{emulateapj}

\newcommand{\mi}{\relax \ifmmode \mu{\mbox m}\else $\mu$m\fi}
\newcommand{\hii}{\relax \ifmmode {\mbox H\,{\scshape ii}}\else H\,{\scshape ii}\fi}
\newcommand{\ha}{\relax \ifmmode {\mbox H}\alpha\else H$\alpha$\fi}
\newcommand{\pa}{\relax \ifmmode {\mbox Pa}\alpha\else Pa$\alpha$\fi}
\newcommand{\hb}{\relax \ifmmode {\mbox H}\beta\else H$\beta$\fi}
\newcommand{\ergs}{\relax \ifmmode {\,\mbox{erg\,s}}^{-1}\else \,\mbox{erg\,s}$^{-1}$\fi}

\begin{document}
\title{On the metallicity dependence of the 24\mi\ luminosity as a star
   formation tracer}
\author{M. Rela\~no\altaffilmark{1}, U. Lisenfeld\altaffilmark{1},
  P. G. P\'erez-Gonz\'alez\altaffilmark{2,3},
  J.M. V\'\i lchez\altaffilmark{4},
   E. Battaner\altaffilmark{1}}

\altaffiltext{1}{Dpto. de F\'\i sica Te\'orica y del Cosmos, 
	      Universidad de Granada, Avda. Fuentenueva s/n, 18071, 
Granada, Spain; mrelano@ugr.es, ute@ugr.es, battaner@ugr.es}
\altaffiltext{2}{Dpto. de Astrof\'\i sica y CC. de la Atm\'osfera, Facultad de CC. F \'\i sicas, Universidad Complutense de Madrid, 28040, Madrid, Spain; pgperez@astrax.fis.ucm.es}
\altaffiltext{3}{Associate Astronomer at Steward Observatory, The 
University of Arizona}
\altaffiltext{4}{Instituto de Astrof\'\i sica de Andaluc\'\i a, CSIC, 
Apartado 3004, 18080, Granada, Spain; jvm@iaa.es}

\begin{abstract}
We investigate the use of the rest-frame 24\mi\ luminosity as an indicator of the star formation rate (SFR)
  in galaxies with different metallicities by comparing it to the (extinction corrected)
  \ha\ luminosity. 
We carry out this analysis in 2 steps: First, we compare the emission
from \hii\ regions in different galaxies with metallicities between 
12+log(O/H) = 8.1 and 8.9.
We find that the 24\mi\ and the 
extinction corrected \ha\ luminosities from individual \hii\ regions follow the same 
correlation for all galaxies, independent of their metallicity. 
Second, the role of metallicity is explored further for the integrated luminosity in a sample of galaxies with metallicities in the range of 12+log(O/H) = 7.2 -- 9.1.
For this sample we compare the 24\mi\ and \ha\ luminosities integrated over the
entire galaxies and find a lack of the 24\mi\ emission for a given \ha\ luminosity for low metallicity objects, likely reflecting a low dust content. 
These results suggest that the 24\mi\ luminosity is a good metallicity independent 
tracer for the SFR in individual \hii\ regions. On the other hand,
metallicity has to be taken into account
when using the 24\mi\ luminosity as a tracer for the SFR of entire galaxies.
\end{abstract}

\section{Introduction}

The total infrared (IR) emission is known to be an optimum tracer of the star formation rate (SFR) 
for highly obscured star-forming regions (Kennicutt
1998; Sanders \& Mirabel 1996). Observationally, the power of the IR (especially the
mid-IR) to trace star formation (SF) has been confirmed using
Infrared Space Observatory data (Genzel \& Cesarsky 2000). Detailed studies
of extended SF along the spiral arms of normal disk
galaxies carried out by Roussel et al. (2001) showed that the SFR can be
parametrized by the luminosity at 7 or 15\mi, and this has been
confirmed by a recent study of 20 spiral and starburst galaxies (Forster Schreiber et al. 2004). Correlations of the total IR luminosity and the luminosity at 6.7, 12 and 15\mi\ are shown in Chary \& Elbaz (2001).
 
After the launch of the Spitzer Space Telescope (Werner et al. 2004), 
new IR wavelength bands have
been proposed to trace the SFR in late-type spiral
galaxies. Comparisons of these new bands (24\mi\ from MIPS, 
Rieke et al. 2004, and 8\mi\ from IRAC, Fazio et al. 2004) with other
typical SFR tracers, such us the \ha\  emission, 
have shown very good correlations that hold over more than
two orders of magnitude in luminosity (Calzetti et al. 2005 (CKB);
P\'erez-Gonz\'alez et al. 2006 (PKG)). The correlation found between the
24\mi\ luminosity and the extinction corrected \ha\ luminosity for the
central \hii\ emitting knots in M51 was later confirmed for the 
 \hii\ regions in M81. In the latter object, the dispersion was
however found to be  higher, which  was explained by the
significant amount of non-obscured SF and by the large uncertainties in the attenuation
estimations.
Recently, Calzetti et al. (2007) carried out a detailed study of the
mid-IR emission as a SFR indicator and concluded that the 24\mi\ emission
shows a good, however non-linear relation with the \pa\ emission.
They have also explored the possible role of the metallicity on this relation. 

Other studies have investigated the relation between the
24\mi\ and the extinction corrected \ha\ luminosities in other types
of galaxies. Alonso-Herrero et al. (2006) obtained the same good
correlation between the 24\mi\ and the extinction-corrected 
\pa\ luminosities of Luminous Infrared Galaxies (LIRGs) 
and Ultraluminous Infrared Galaxies (ULIRGs), which makes the
relation applicable over nearly five orders of magnitude in luminosity. Wu et
al. (2005) also found a good correlation between the integrated 24\mi\
and 8\mi\ luminosities and the \ha\ luminosity in a sample of star-forming
galaxies, but they obtained a change in the slope for the dwarf
galaxies of their sample. 
Cannon et al. (2005, 2006ab) studied \ha\ and Spitzer data of
 the dwarf galaxies IC~2574, NGC~1705 and 
NGC~6822, respectively. They found values for the 24\mi\ luminosities of the \hii\ 
regions 3-5 times lower than expected from the \ha\ luminosity when
applying the relation for M51 of CKB.
The aim of this Letter is to investigate the reason of the differences found 
in the references above,
and in particular to study the role played by the metallicity in the 24\mi-\ha\ relation.
 
\section{Galaxy sample and Data Analysis}

Our study is based, apart from data of the literature (see below)
on our own analysis of the nearby dwarf galaxies NGC~1569 and
NGC~4214, for which Spitzer MIPS images at 24\mi\ are
available.

The optical data of the dwarf galaxies NGC~1569 and NGC~4214 were
taken from the HST data archive. The data analysis for the \ha\ and \hb\ 
images of NGC~1569 is explained in Rela\~no et
al. (2006); for NGC~4214, we obtained \ha\ and \hb\ images following the procedure explained in MacKenty et al. (2000). 
The \ha\ fluxes of the most luminous \hii\ regions 
in NGC~1569 coincide within  10\% with the values given by Waller (1991). For 
NGC~4214 the differences with respect to the fluxes of MacKenty et al. (2000) are less than 16\%. 
The MIPS 24\mi\ images of these galaxies 
were taken from the Spitzer Data Archive and reduced using the MIPS Data 
Analysis Tool (Gordon et al. 2005). The calibration uncertainties amount to $\sim$4\% (Engelbracht et al. 2007). 

In order to make the \ha\ and IR photometric analysis 
in NGC~1569 and NGC~4214, all images were convolved to the
resolution of the MIPS 24\mi\ image, using the
semi-empirical PSF for a 75~K blackbody\footnote{
http://dirty.as.arizona.edu/$\sim$kgordon/mips/conv$_{-}$psfs/conv$_{-}$psfs.html}. 
The \ha\ and 24\mi\ images have the same appearance: intense knots
observed in \ha\ match with bright emission at 24\mi. For NGC~1569, we selected 6 apertures 
corresponding to the most luminous \hii\ regions catalogued in Waller (1991). Two of these 
apertures contain, additionally, a nearby 
less luminous \hii\ region which we could not resolve separately due to the
coarse  spatial resolution ($\sim$6\arcsec)  of our images. 
In the case of NGC~4214, the apertures selected correspond to the 
largest \hii\ complexes defined in Table~2 of MacKenty et al. (2000). 
In both objects, we selected the aperture size (between 11\arcsec\ and 33\arcsec, 
corresponding to  200-300~pc) individually to match the size of
each \hii\ region. Finally, aperture corrections derived from the theoretical PSF 
at 24\mi\ were applied to the \ha\ and 24\mi\ luminosities.

The Balmer extinction of the \hii\ regions in NGC~1569 was obtained from the corresponding  integrated \ha\ and \hb\ fluxes and following Caplan \& Deharveng (1986). The extinction values for each \hii\ region agree with those in the extinction map shown in Rela\~no et al. (2006). 
Extinction values for each \hii\ region in NGC~4214 were taken from MacKenty et al. (2000), who 
applied a foreground dust screen model and used bidimensional spectroscopy
studies from Ma\'\i z-Apell\'aniz et al. (1998).  

We also used 24\mi\ and \ha\ luminosities integrated over the entire galaxies (see Table~1 for the values and their references).
For the \ha\ luminosity of M51 we applied an extinction correction of $\rm A_{\scriptsize\ha}$=1~mag, which is an intermediate 
value to those given in CKB for the central part of the galaxy and the 
values reported in Bresolin et al. (2004) for the \hii\ regions in the outer part of the galaxy. 
We estimate the uncertainty in the extinction to be $\sim 1$ mag, resulting in
an uncertainty in the extinction corrected \ha\ luminosity of 0.4~dex. 
For M81, the observed \ha\ flux  
was corrected with an extinction of $\rm A_{\scriptsize\ha}$=0.3~mag, derived using an average 
value of the interstellar reddening map of M81 (Kong et al. 2000) 
and the extinction law of 
Draine (2003) with $\rm R_V=3.1$. From the spatial variations of 
the reddening map of this galaxy we estimate an uncertainty of 0.3~mag, resulting in an
uncertainty of the extinction corrected \ha\ luminosity of 0.12~dex. 
For NGC~1569 we used the mean value of the total extinction given by 
Devost et al. (1997), $\rm A_{\scriptsize\ha}$=1.78~mag, 
in agreement with Rela\~no et al. (2006); 
and for NGC~4214, $\rm A_{\scriptsize\ha}$=0.3~mag, derived from maps of the Balmer ratio 
shown in Ma\'\i z-Apell\'aniz et al. (1998) and applied by 
MacKenty et al. (2000). For both galaxies we estimate the uncertainty to be 0.3~mag,
yielding  in an
uncertainty of the extinction corrected \ha\ luminosity of 0.12~dex.
The \ha\ luminosities of NGC~1705, NGC~6822 and IC~2574  were not
corrected for internal extinction which was however shown to be small 
(Cannon et al. 2005, 2006ab). 

Finally, we have included a
sample of dwarf galaxies covering a wide range of metallicities. The sample is
composed of the dwarfs observed by MIPS (Engelbracht et al. 2005),
counting with published Galactic extinction corrected \ha\ fluxes (Gil de Paz et al. 2003). 
Given the
low metallicity of these dwarf galaxies, internal extinction should be small and is 
not expected to affect  the 
conclusions of our study. We also added galaxies from 
the LIRG and ULIRG sample of Alonso-Herrero et al. (2006) with metallicity values available in 
the literature. We eliminated two galaxies (IC~860 and NGC~7469) of this sample 
showing high IRAS infrared 
emission from their nucleus, possibly due to an
active galactic nucleus.
The combined galaxy sample, extinction corrected (as described above) \ha\ and
 24\mi\ luminosities together with 
the metallicities and distances are listed in
Table~1. 

\begin{deluxetable}{llllllll}
\tabletypesize{\tiny}
\tablecaption{Galaxy sample and metallicities}
\tablewidth{0pt}
\tablehead{
\colhead{\tiny{Galaxy}} & \colhead{\tiny{$\rm log(L_{\tiny{24}})$}} & \colhead{\tiny{Ref}}& \colhead{\tiny{$\rm log(L_{\tiny{\ha}}^{\tiny{corr}})$}} & \colhead{\tiny{Ref}} & \colhead{\tiny{$\rm Z$}} & \colhead{\tiny{Ref}} & \colhead{\tiny{$\rm D$}} \\
\colhead{} & \colhead{\tiny{(\ergs)}} & \colhead{}& \colhead{\tiny{(\ergs)}} & \colhead{} & \colhead{} & \colhead{} & \colhead{\tiny{(Mpc)}}}
\startdata
\tiny{I~Zw~18}         & \tiny{40.20} & \tiny{2,3} & \tiny{39.83} & \tiny{1} & \tiny{7.2} & \tiny{7} & \tiny{12.6} \\
\tiny{HS0822+3542}     & \tiny{39.62} & \tiny{3} & \tiny{38.95} & \tiny{1} & \tiny{7.4} & \tiny{8} & \tiny{10.1} \\
\tiny{Tol~65}          & \tiny{41.46} & \tiny{3} & \tiny{40.59} & \tiny{1} & \tiny{7.6} &\tiny{7} & \tiny{36.0} \\
\tiny{VII~Zw~403}      & \tiny{39.99} & \tiny{3} & \tiny{39.21} & \tiny{1} & \tiny{7.7} & \tiny{9} & \tiny{4.8} \\
\tiny{II~Zw~70}        & \tiny{41.64} & \tiny{2} & \tiny{40.54} & \tiny{1} & \tiny{7.8} & \tiny{10} &\tiny{18.7} \\ 
\tiny{NGC~4861}        & \tiny{41.82} & \tiny{3} & \tiny{41.03} & \tiny{1} & \tiny{7.9} & \tiny{11} & \tiny{12.6} \\
\tiny{Mrk~1450}        & \tiny{41.19} & \tiny{3} & \tiny{40.04} & \tiny{1} & \tiny{8.0} & \tiny{30} & \tiny{14.7} \\
\tiny{I~Zw~40}        & \tiny{42.33} & \tiny{3} & \tiny{41.25} & \tiny{1} & \tiny{8.1} &\tiny{7} &\tiny{9.8} \\
\tiny{NGC~6822}        & \tiny{39.96}& \tiny{23} & \tiny{39.30} & \tiny{23} & \tiny{8.1} & \tiny{24} & \tiny{0.49} \\
\tiny{IC~2574}        & \tiny{40.73} & \tiny{25}& \tiny{39.98} &\tiny{26} & \tiny{8.15} & \tiny{27} & \tiny{4.0} \\
\tiny{NGC~1705}       & \tiny{40.30} & \tiny{28} & \tiny{39.90} & \tiny{28} &\tiny{8.21} &\tiny{29} &\tiny{5.1} \\
\tiny{NGC~4670}        & \tiny{41.87} & \tiny{3} & \tiny{40.79} &\tiny{1} & \tiny{8.2} & \tiny{13} & \tiny{15.3} \\
\tiny{NGC~1569}        & \tiny{41.73} & \tiny{2} & \tiny{40.80} &\tiny{2} & \tiny{8.2} & \tiny{31} & \tiny{2.2} \\
\tiny{NGC~4214}        & \tiny{41.29} & \tiny{2} & \tiny{40.27} &\tiny{2} &\tiny{8.2} & \tiny{7} &\tiny{2.9} \\
\tiny{IC~4518A}        & \tiny{43.87} & \tiny{4} & \tiny{42.21} & \tiny{4} & \tiny{8.6} &\tiny{14} &\tiny{69.9} \\
\tiny{IC~4518B}        & \tiny{43.33} & \tiny{4} & \tiny{41.66} & \tiny{4} & \tiny{8.6} & \tiny{14} & \tiny{69.9} \\
\tiny{NGC~5135}        & \tiny{43.92} & \tiny{4}& \tiny{42.20} & \tiny{4} &\tiny{8.7} & \tiny{15} & \tiny{52.2} \\
\tiny{NGC~2537}        & \tiny{41.23} & \tiny{3}& \tiny{40.14} & \tiny{1} & \tiny{8.7} &\tiny{12} & \tiny{6.9} \\
\tiny{M81}            & \tiny{41.99} & \tiny{5} & \tiny{40.85} &\tiny{22,2} & \tiny{8.7} & \tiny{20} & \tiny{3.6} \\
\tiny{MCG-02-33-098}   & \tiny{44.04} & \tiny{4} & \tiny{42.26} & \tiny{4} & \tiny{8.7} &\tiny{16} & \tiny{72.5} \\
\tiny{NGC~7771}      & \tiny{43.96} & \tiny{4} & \tiny{42.59} &
\tiny{4} & \tiny{8.8} & \tiny{16} &\tiny{57.1} \\
\tiny{NGC~3690}        & \tiny{44.46} & \tiny{4}& \tiny{42.78}& \tiny{4} & \tiny{8.8} &\tiny{17} & \tiny{47.7} \\
\tiny{NGC~7130}       & \tiny{44.08} & \tiny{4} & \tiny{42.24} & \tiny{4} & \tiny{8.8} &\tiny{16} &\tiny{66.0} \\
\tiny{UGC~3351}        & \tiny{43.61} & \tiny{4}& \tiny{42.34} &\tiny{4} & \tiny{8.8} & \tiny{18} & \tiny{60.9} \\	
\tiny{NGC~3256}        & \tiny{44.41} & \tiny{4} & \tiny{42.55} &\tiny{4} & \tiny{8.8} & \tiny{19} & \tiny{35.4} \\
\tiny{IC~4687}        & \tiny{44.27} & \tiny{4} & \tiny{42.57} &\tiny{4} & \tiny{8.8} &\tiny{15} & \tiny{74.1} \\
\tiny{IC~5179}        & \tiny{43.83} & \tiny{4} & \tiny{42.41} & \tiny{4} & \tiny{8.9} & \tiny{16} &\tiny{46.7} \\
\tiny{M51}          & \tiny{43.09} & \tiny{25}& \tiny{41.72} & 
\tiny{22,2} & \tiny{8.9} & \tiny{20} & \tiny{8.2} \\
\tiny{IRAS17138-1017}  & \tiny{44.20}& \tiny{4} & \tiny{42.42} &\tiny{4} & \tiny{8.9} & \tiny{15} & \tiny{75.8} \\
\tiny{NGC~2369}       & \tiny{43.74}& \tiny{4} & \tiny{42.10} &
\tiny{4} & \tiny{8.9} &\tiny{16} & \tiny{44.0} \\
\tiny{NGC~6701}        & \tiny{43.74} & \tiny{4}& \tiny{42.13} & \tiny{4} & \tiny{8.9} & \tiny{16} & \tiny{56.6} \\
\tiny{NGC~633}        & \tiny{43.63}& \tiny{4}& \tiny{41.85} & 
\tiny{4} & \tiny{8.9} & \tiny{15} & \tiny{67.9} \\
\tiny{NGC~7591}       & \tiny{43.85} & \tiny{4} & \tiny{42.03} & \tiny{4} & \tiny{8.9} & \tiny{16} & \tiny{65.5} \\
\tiny{NGC~5653}       & \tiny{43.73} & \tiny{4} & \tiny{42.21} &
\tiny{4} & \tiny{8.9} & \tiny{16} & \tiny{54.9} \\
\tiny{NGC~23}         & \tiny{43.77} & \tiny{4} & \tiny{42.40} & \tiny{4} & \tiny{8.9} & \tiny{16} & \tiny{59.6} \\
\tiny{IC~4734}        & \tiny{43.91}& \tiny{4} & \tiny{42.13} & \tiny{4} & \tiny{9.0} & \tiny{14} & \tiny{68.6} \\
\tiny{NGC~5936}       & \tiny{43.85}&\tiny{4} & \tiny{42.14} &
\tiny{4} & \tiny{9.0} & \tiny{16} & \tiny{60.8} \\
\tiny{NGC~5734}       & \tiny{43.52} & \tiny{4} & \tiny{41.99} &\tiny{4} & \tiny{9.0} & \tiny{16} & \tiny{59.3} \\
\tiny{NGC~3110}       & \tiny{43.90} &\tiny{4} & \tiny{42.50} & \tiny{4} & \tiny{9.0} & \tiny{16} &\tiny{73.5} \\
\tiny{ESO320-G030}   & \tiny{43.62}& \tiny{4} & \tiny{41.91} & 
\tiny{4} & \tiny{9.0} & \tiny{21} & \tiny{37.7} \\
\tiny{NGC~2388}      & \tiny{43.93} & \tiny{4} & \tiny{42.23} &
\tiny{4} & \tiny{9.1} &\tiny{16} & \tiny{57.8} \\
\enddata
\tablecomments{\tiny{
Z is the oxygen abundance, 12+log(O/H). We have checked that the derivation of these values is consistent for this sample. Ref is the code for the reference:
(1) Gil de Paz et al. 2003; (2) This paper; 
(3) Engelbracht et al. 2005; (4) Alonso-Herrero et al. 2006; 
(5) P\'erez-Gonz\'alez et al. 2006 (PKG); (6) Calzetti et al. 2005 (CKB);
(7) Kobulnicky \& Skillman 1996; (8) Kniazev et al. 2000; 
(9) Izotov et al. 1997; (10) Shi et al. 2005; (11) Kobulnicky et al. 1999;
(12) Guseva et al. 2000; (13) Heckman et al. 1998;  
(14) Corbett et al. 2003; (15) Kewley et al. 2001; (16) Veilleux et al. 1995;
(17) Armus et al. 1989 (18) Baan et al. 1998; (19) Sekiguchi \& Wolstencroft 1993;
(20) Pilyugin et al. 2004; (21) van den Broek et al. 1991; (22) Greenawalt et al. 1998.(23) Cannon et al. 2006b; (24) Lee et al. 2006; (25) Dale et al. 2005; (26) Miller \& Hodge 1994; (27) Miller \& Hodge 1996; (28) Cannon et al. 2006a; (29) Lee \& Skillman 2004; (30) Izotov et al. 1994; (31) Kobulnicky \& Skillman 1997.}}
\end{deluxetable}

\section{Results}
\begin{figure}
\plotone{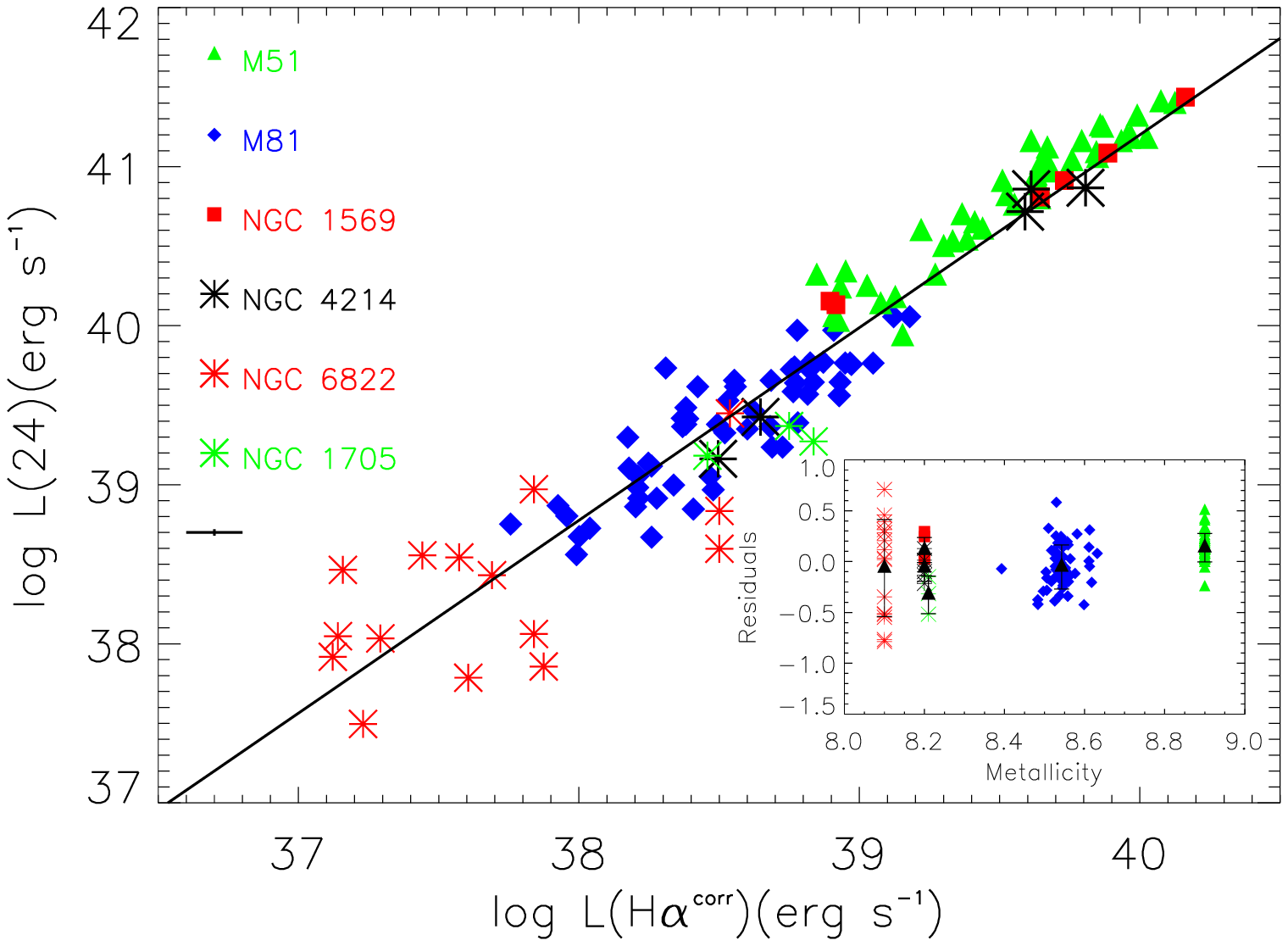}
\plotone{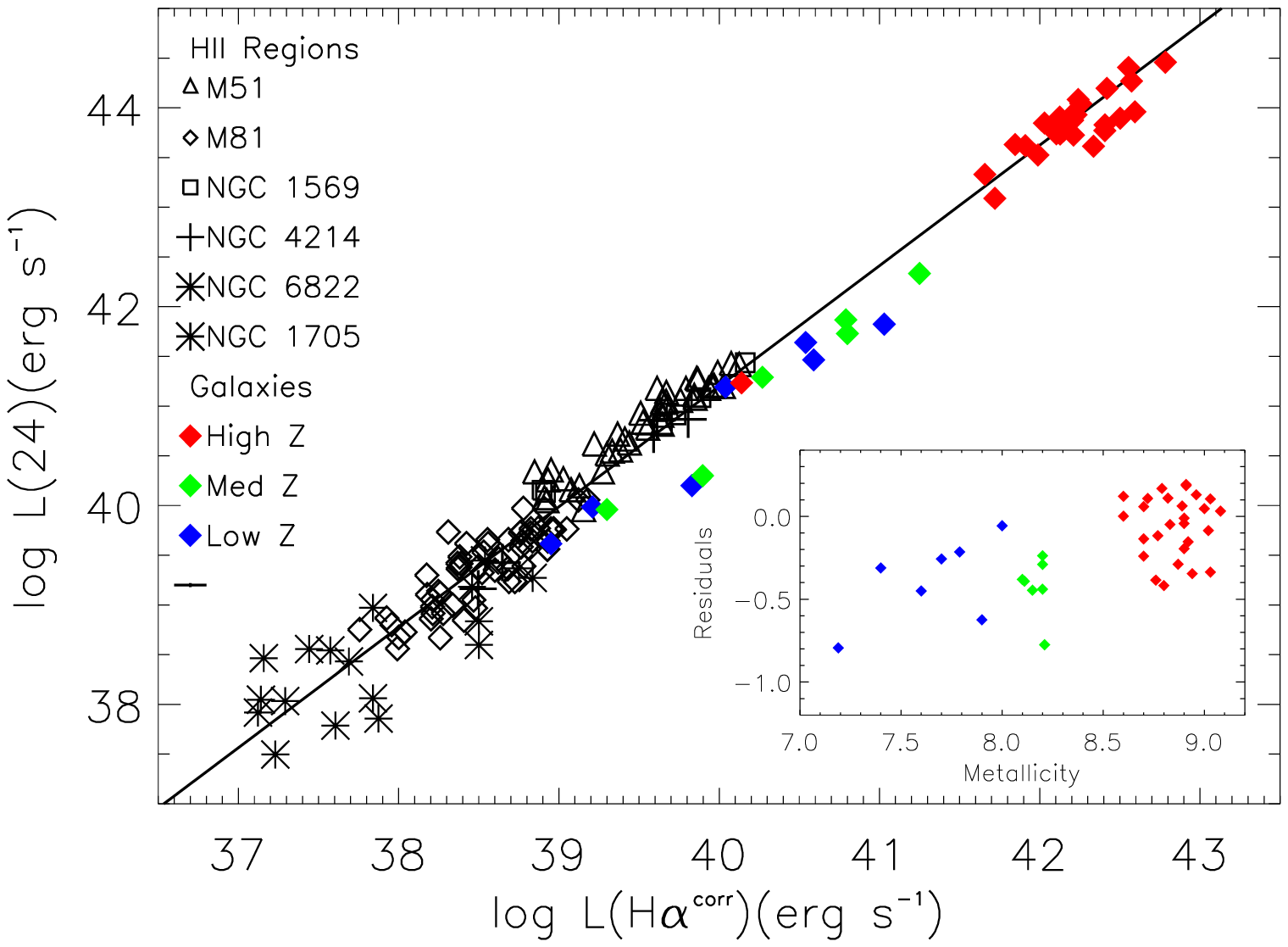}
\caption{{\it Top:} 24\mi\ luminosity as a function of the extinction corrected \ha\ 
luminosity for a sample of \hii\ regions in M51, M81, NGC~1569, NGC~4214, NGC~6822 and 
NGC~1705. The solid line shows the linear fit to the \hii\ regions of M51, M81 and the ULIRGs from Alonso-Herrero et al. (2006) (see Eq.~1).
Typical error bars are shown in the lower left corner of the plot: 
they account for uncertainties in the calibration (4\% for 24\mi\ flux 
(Engelbracht et al. 2007), $\sim$15\% for \ha\ flux and $\sim$0.2mag for the extinctions 
(Gil de Paz et al. 2003; Rela\~no et al. 2006)).
In the inner panel we show the residuals of the
24\mi\ luminosity (see text) {\it versus} the metallicity. The black triangles 
represent the mean value of the residuals for each 
galaxy. 
{\it Bottom:} The same plot as above but 
including the integrated luminosities of the galaxies in Table~1. The solid line is
the linear fit shown in Eq.~1. 
We use different colors for galaxies with different metallicities: Z$\leq$8.0 (blue), 
8.0$<$Z$<$8.5 (green) and Z$\geq$8.5 (red). 
In the inner panel we show the residuals of the 24\mi\ luminosity 
with respect to linear fit (Eq.~1) {\it versus} the metallicity.} 
\label{lumi}
\end{figure}

In Fig.~\ref{lumi} (top panel) we compare the extinction corrected \ha\ luminosities
and the 24\mi\ luminosities for the \hii\ regions in NGC~1569 and NGC~4214 with those published for 
M51 (CKB), M81  (PKG), NGC~1705 (Cannon et al. 2006a) and NGC~6822 (Cannon et al. 2006b). 
The data points of the three low-metallicity galaxies (NGC~1569, NGC~4214 and NGC~6822)
follow closely the same distribution as the combined data set of the higher metallicity
galaxies M51 and M81 (green triangles and blue diamonds, respectively). 
The \hii\ regions of NGC~6822, representing the lowest
luminosities, show  a larger scatter
than the rest of the \hii\ regions.  
A possible reason could be the fact that the surface brightness of 
the \hii\ regions in NGC~6822 is 1-2 orders of magnitude lower than the 
surface brightness of the \hii\ regions in the other objects.
In order to search for differences as a function of metallicity we
have derived a linear fit  including the \hii\ regions of 
the high metallicity galaxies (M51 and M81) and the ULIRG sample of Alonso-Herrero et al. (2006)
yielding: 
\begin{equation}
\rm logL(24)=(-7.28\pm 0.52)+(1.21\pm 0.01)\times logL(\ha^{corr})
\end{equation}
The slope of this fit is similar to the linear fit obtained by Calzetti et al. (2007) for 
the high metallicity data points of their sample, derived using their Eqs.~6 and~9.
In the inner plot of Fig.~\ref{lumi} (top panel) we show the residuals of the
24\mi\ luminosity (i.e. the difference between the logarithm of the measured 
24\mi\ luminosity and the logarithm of the expected luminosity from  
the linear fit given in Eq.~1) {\it versus} the metallicity of each \hii\ region.
For M81 and M51 we estimate the metallicity of the \hii\ regions using 
the metallicity gradients of each galaxy derived by Pilyugin et al (2004). 
For the \hii\ regions in M51 we derive a metallicity variation of $<$0.1dex for the radial range of their galactocentric radius, and therefore we adopt the 
central metallicity value for all of them.
For the rest of the galaxies in Fig.~\ref{lumi} (top panel), 
there is no appreciable metallicity gradients (see references in Table~1).
No trend with metallicity is visible, with 
the  mean of the residuals for the \hii\ regions of each galaxy being practically 
constant over the whole metallicity range.
Thus we conclude that, within the metallicity range investigated here, 
the relation between the 24\mi\ and \ha\ luminosities of \hii\ regions
shows no dependence on metallicity.

The situation changes when the integrated galaxy luminosities are considered.
In Fig.~\ref{lumi} (bottom panel) we compare the data for \hii\ regions with the integrated 
luminosities of the galaxy sample of Table~1, which includes dwarf galaxies with low metallicities 
and (U)LIRGs with 
high metallicities. For these additional galaxies, data for individual \hii\ regions is not available. We find a trend that low metallicity galaxies fall  below the linear fit shown in Eq.~1, whereas the high metallicity galaxies follow it. In the inner plot of this figure we show again the residuals of the 24\mi\ luminosity (with respect to the fit of Eq.~1) 
{\it versus} the metallicity of the galaxy.
A trend of lower metallicity galaxies to have a lower ratio of measured-to-expected values is visible, with a correlation coefficient of 0.63. We expect that 
the uncertainties in the extinction correction of the \ha\ fluxes 
will not change this trend  for two reasons: The low metallicity dwarf galaxies, uncorrected for
internal extinction, would be located even further away from the regression fit 
if we had applied an internal extinction correction, which would further 
emphasize the observed trend. The 
rest of the galaxies (except M81 and M51) show only small uncertainties in the adopted extinction values. The higher uncertainties in
the case of M81 and M51 is not able to change the general trend 
observed in the bottom panel of Fig.~\ref{lumi}.

\section{Discussion}

It is surprising that the relation between
the 24\mi\ and \ha\ luminosities for individual \hii\ regions does not show any dependence on metallicities
for the range investigated here (galaxies with metallicities between 12+log(O/H) = 8.1 and 8.9).
A possible reason might be the existence of a lower threshold for the accumulation 
of dust in HII regions in order to support SFRs as large as the ones measured in 
our HII regions ($\sim 10^{-4}-10^{-1}$ M$_\odot$ yr$^{-1}$).
This would result in the dust content of 
HII regions being independent of the global metallicity of the entire galaxy. 
Support for this hypothesis comes from 
a detailed study of the Balmer decrement in 
the major  \hii\ regions of NGC~1569 (Rela\~no et al. 2006), where locally 
a high intrinsic  extinction (A$_V = 0.8$mag) was found, 
in spite of the low metallicity of this galaxy.
Further studies for higher and lower metallicity galaxies are needed to show how
far the universality of the relation between the 24\mi\ and \ha\ luminosity extends.

The situation is very different for the ISM outside the \hii\ region, the diffuse ISM,
which we  take into account when considering the 24\mi\ and \ha\ emission
integrated over the entire galaxy. 
Here, the 24\mi\ emission depends directly on the dust content and opacity and 
hence on the metallicity. Recently, Calzetti et al. (2007) studied the relation between the 24\mi\ and the extinction 
corrected \pa\ surface densities of the star forming regions in a sample of nearby galaxies. They found a slight trend for \hii\ regions of low-metallicity galaxies to have a lower 24\mi\ emission for a given \pa\ surface density than higher metallicity regions. Their use of a fixed aperture size 
for the \hii\ regions in all galaxies migh have included some diffuse emission, especially
for distant galaxies. This would explain the similarity of their results and ours
for the integrated emissions.
A deeper analysis of the results presented in this letter, including a larger sample of \hii\ regions and separating clearly the diffuse emission from the emission coming from the \hii\ regions is needed (Rela\~no et al. in prep.).

\section{Summary and conclusions}

We have studied the role of the metallicity in the use of the 24\mi\ 
luminosity as a SFR indicator by analyzing the data for a sample
of dwarfs and spirals covering a wide range of metallicities. We 
found that the extinction corrected \ha\ and the 
24\mi\ luminosities correlate tightly for all \hii\ regions,
independently of the global metallicity of the galaxy. This demonstrates
that the 24\mi\ luminosity is a good tracer of the {\it local} SFR,
independent of the metallicity. This is not the case when considering
the integrated emission of galaxies. In this case, metal--rich galaxies
present a higher 24\mi\  luminosity for a given \ha\ luminosity than low
metallicity galaxies. Our results indicate that the 24\mi\ luminosity
can be used as a SFR tracer when taking into account: (i) whether the
emission from \hii\ regions or the integrated emission from entire
galaxies is considered, and (ii) the metallicity of the galaxy.

\acknowledgments

We would like to thank the anonymous referee and Almudena Zurita for useful suggestions that improve the final version of the manuscript.
This work has been supported by the Spanish Ministry of Education and Science
within the {\it PNAYA} via projects AYA2004-08251-C02-00, and ESP2003-00915. PGPG 
acknowledges support from
the Ram\'on y Cajal Fellowship Program and the project AYA
2006--02358.


\begin{thebibliography}{}
\bibitem[]{437} Alonso-Herrero, A., Rieke, G., Rieke, M. et al. 2006, \apj, 650, 835
\bibitem[]{439} Armus, L., Heckman, T. M., Miley, G. K., 1989, \apj, 347, 727
\bibitem[]{440} Baan, W., A., Salzer, J. J., LeWinter, R. D. 1998, \apj, 509, 633
\bibitem[]{441} Bresolin, F., Garnett, D. R., Kennicutt, R. C. Jr. 2004, \apj, 615, 228
\bibitem[]{442} Calzetti, D., Kennicutt, R. C. Jr., Bianchi, L. et
  al. 2005, \apj, 633, 871 (CKB)
\bibitem[]{442} Calzetti, D., Kennicutt, R. C. Jr., Engelbracht, C. W. et al. 2007, astro-ph:0705.3377
\bibitem[]{444} Cannon, J. M., Walter, F., Bendo, G. et al. 2005, \apjl, 630, 37
\bibitem[]{445} Cannon, J. M., Smith, J-D., T., Walter, F. et al. 2006a,
  \apj, 647, 293
\bibitem[]{447} Cannon, J. M., Walter, F., Armus, L. 2006b, \apj, 652, 1170
\bibitem[]{448} Caplan, J., Deharveng, L. 1986, \aap, 155, 297
\bibitem[]{449} Chary, R., Elbaz, D. 2001, \apj, 556, 562
\bibitem[]{450} Corbett, E. A., Kewley, L., Appleton, P. N., Charmandaris, V., Dopita, M. A., Heisler, C. A., Norris, R. P., Zezas, A., Marston, A. 2003, \apj, 583, 670
\bibitem[]{451} Dale, D. A., Bendo, G. J., Engelbracht, C. W. et al. 2005, \apj, 633, 857 
\bibitem[]{452} Devost, D., Roy, J. R., Drissen,
   L. 1997, \apj, 482, 765
\bibitem[]{454} Draine, B. T. 2003, \araa, 41, 241
\bibitem[]{455} Engelbracht, C. W., Gordon, K. D., Rieke, G. H., Werner,
  M. W., Dale, D. A., Latter, W. B. 2005, \apjl, 628, 29
\bibitem[]{457} Engelbracht, C. W. et al. 2007, astro-ph:0704.2195v1
\bibitem[]{458} Fazio, G. G., Hora, J. L., Allen, L. E. et al. 2004, ApJS, 154, 10 
\bibitem[]{460} Forster Schreiber, N. M., Roussel, H., Sauvage, M.,
  Charmandaris, V. 2004, \aap, 419, 501
\bibitem[]{462} Genzel, R., Cesarsky, C. J. 2000, \araa, 38, 761
\bibitem[]{463} Gil de Paz, A., Madore, B. F., Pevunova, O. 2003, \apjs,
  147, 29
\bibitem[]{465} Gordon,  et al. 2005, PASP, 117, 503 
\bibitem[]{466} Greenawalt, B., Walterbos, R. A. M., Thilker, D., Hoopes, C. G. 1998, \apj, 506, 135 
\bibitem[]{467} Guseva, N. G., Izotov, Yu. I., Thuan, T. X. 2000, \apj, 531, 776
\bibitem[]{468} Heckman, T. M., Robert, C., Leitherer, C., Garnett, D. R., van der Rydt, F. 1998,\apj, 503, 646
\bibitem[]{470} Izotov, Yu. I., Thuan, T. X., Lipovetsky, V. A. 1994, \apj, 435, 647 
\bibitem[]{471} Izotov, Yu. I., Thuan, T. X., Lipovetsky, V. A. 1997, \apjs, 108, 1
\bibitem[]{472} Kniazev, A. Yu., et al. 2000, \aap, 357, 101
\bibitem[]{473} Kennicutt, R. C. Jr. 1998, \araa, 36, 189
\bibitem[]{474} Kennicutt, R. C. Jr. et al. \pasp, 115, 928 
\bibitem[]{475} Kewley, L. J., Heisler, C. A., Dopita, M. A., Lumsden, S. 2001, \apjs, 132, 37
\bibitem[]{476} Kobulnicky, H. A., Skillman, E. D. 1996, \apj, 471, 211
\bibitem[]{477} Kobulnicky, H. A., Skillman, E. D. 1997, \apj, 489, 636
\bibitem[]{478} Kobulnicky, H. A., Kennicutt, R. C. Jr., Pizagno, J. L. 1999, \apj, 514, 544 
\bibitem[]{479} Kong, X. et al. 2000, \aj, 119, 2745
\bibitem[]{480} MacKenty, J. W., Ma\'\i z-Apell\'aniz, J., Pickens, C. E.,
  Norman, C. A., Walborn, N. R. 2000, \apj, 120, 3007
\bibitem[]{482} Ma\'\i z-Apell\'aniz, J., Mas-Hesse, J. M.,
  Mu\~noz-Tu\~non, C., V\'\i lchez, J. M., Casta\~neda, H. O. 1998,
  \aap, 329, 409
\bibitem[]{487} Miller, B. W., Hodge, P. 1994, \apj, 427, 656
\bibitem[]{488} Miller, B. W., Hodge, P. 1996, \apj, 458, 467
\bibitem[]{490} Lee, H., Skillman, E. D., Venn, K. 2006, \apj, 642, 813 
\bibitem[]{491} Lee, H., Skillman, E. D. 2004, \apj, 614, 698
\bibitem[]{492} P\'erez-Gonz\'alez, P. G., Kennicutt, R. C. Jr., Gordon,
  K. et al. 2006, \apj, 648, 987 (PKG)
\bibitem[]{494} Pilyugin, L. S., V\'\i lchez, J. M., Contini, T. 2004, \aap, 425, 849
\bibitem[]{495} Rela\~no, M., Lisenfeld, U., V\'\i lchez, J. M.,
  Battaner, E. 2006, \aap, 452, 413
\bibitem[]{497} Rieke, G. H., Young, E. T., Engelbracht, C. W. et al. 2004, ApJS, 154, 25
\bibitem[]{498} Roussel, H., Sauvage, M., Vigroux, L., Bosma, A. 2001,
  \aap, 372, 427
\bibitem[]{501} Sekiguchi, K., Wolstencroft, R. D. 1993, \mnras, 263, 349
\bibitem[]{502} Shi, F., Kong, C. Li, Cheng, F. Z. 2005, \aap, 437, 849
\bibitem[]{503} van den Broek, A. C. et al. 1991, A\&AS, 91, 61
\bibitem[]{504} Veilleux, S., Kim, D.,-C., Sanders, D. B., Mazzarella, J. M., Soifer, B. T. 1995, \apjs, 98, 171
\bibitem[]{505} Waller, W. H., 1991, \apj, 370, 144
\bibitem[]{506} Werner, M. W., Roellig, T. L., Low, F. J. et al. 2004, ApJS, 154, 1
\bibitem[]{507} Wu, H., Cao, C., Hao, C-N. 2005, \apjl, 632, 79
\end{thebibliography}
\end{document}